\documentstyle[12pt,aaspp4]{article}
\begin{document}

\title{A TWO-TEMPERATURE MODEL OF THE INTRACLUSTER MEDIUM}
\author{{\sc Motokazu Takizawa}}
\affil{Department of Astronomy, Faculty of Science, Kyoto University, 
       Sakyo-ku, Kyoto 606-8502, JAPAN}
\authoremail{takizawa@kusastro.kyoto-u.ac.jp}

\begin{abstract}
 We investigate evolution of the intracluster medium (ICM), considering
 the relaxation process between the ions and electrons. According to the
 standard scenario of structure formation, ICM is heated by the shock in
 the accretion flow to the gravitational potential well of the
 dark halo. The shock primarily heats the ions because the kinetic energy
 of an ion entering the shock is larger than that of an electron 
 by the ratio of masses. Then the electrons and 
 ions exchange the energy through coulomb collisions and reach the
 equilibrium. From simple order estimation we find that
 the region where the electron temperature is considerably lower than
 the ion temperature spreads out on a Mpc scale.

 We then calculate the ion and electron temperature profiles by
 combining the adiabatic model of two-temperature plasma by 
 Fox \& Loeb (1997) with spherically symmetric N-body and hydrodynamic
 simulations based on three different cosmological models.
 It is found that the electron temperature is about a half of the mean 
 temperature at radii $\sim$ 1 Mpc. This could lead to an about 50 \%
 underestimation in the total mass contained within $\sim$ 1 Mpc when the
 electron temperature profiles are used. The polytropic indices of the
 electron temperature profiles are $\simeq 1.5$ whereas those of mean 
 temperature $\simeq 1.3$ for $r \geq 1$ Mpc. This result is consistent 
 both with the X-ray  observations on electron temperature profiles and 
 with some theoretical and numerical predictions about mean temperature 
 profiles.
\end{abstract}


\section{INTRODUCTION}
Clusters of galaxies (CG) are the largest virialized objects in the
universe, which contain collisionless particles, galaxies and dark
matter, and a diffuse gas component. The gas component is called
intracluster medium (ICM). The ICM is the plasma with temperature of
about $10^8$ K, thus emitting X-ray mainly through the thermal
bremsstrahlung of the electrons (Sarazin 1988).

Revealing temperature profiles of ICM is an important problem.
In theoretical work concerning CG, isothermality of ICM is often
assumed. In particular, the isothermal $\beta$ model (Cavaliere \&
Fusco-Femiano 1976) is adopted in the conventional mass determination
through X-ray observations, in the estimation of the Hubble constant
($H_0$) through Sunyaev-Zel'dovich (SZ) effect (Sunyaev \&
Zel'dovich 1972; Birkinshaw, Hughes, \& Arnaud 1991), etc. However,
deviation from a flat temperature profile becomes important at lower
densities in the mass estimation using X-ray data (Evrard, Metzler, \&
Navarro 1996; Schindler 1996). Also in the $H_0$ estimation through SZ
effect the non-isothermality is one important source of errors (Inagaki,
Suginohara, \& Suto 1995; Yoshikawa, Itoh, \& Suto 1998). Therefore,
accurate temperature profiles of ICM are required to improve such
methods.

On the other hand, temperature maps of ICM provide us with useful
information about CG. In merging clusters, characteristic temperature
structures are expected to occur through shock heating and adiabatic
compression. Some numerical simulations are especially focused on this
problem (Schindler \& M\"{u}ller 1993; Ishizaka \& Mineshige 1996;
Roettiger, Loken, \& Burns 1997; Ishizaka 1998; 
Roettiger, Stone, \& Mushotzky 1998). 
In particular, off-center collisions are investigated by Ricker (1998).
Comparing these results with the X-ray observational data
, we can guess in what phase the merging clusters are
(Fujita et al. 1996; Honda et al. 1996; Churazov et al. 1998; 
Donnelly et al. 1998; Davis \& White 1998).

Recent X-ray observations with ASCA and ROSAT reveal the radial electron
temperature distribution of ICM (Markevitch et al. 1996; Markevitch 1996; 
Markevitch, Sarazin, \& Irwin 1996; Markevitch et al. 1997). 
In some clusters the observed
electron temperature gradients correspond to the polytropic index, 
$\gamma (P \propto \rho^{\gamma} )$, of $\gamma = 1.5$ or even more
(Markevitch 1996). On the other hand, $\gamma \simeq 1.2$ is expected
for plasma mean temperature distribution derived by the self similar
solution of Bertschinger (1985) and some numerical simulations 
(Evrard 1990; Katz \& White 1993; Navarro, Frenk, \& White 1995; 
Eke, Navarro \& Frenk 1997; Takizawa \& Mineshige 1998; 
Bryan \& Norman 1998). There is a discrepancy between the observed
values and the theoretical expectations.

To explain the discrepancy between the observed electron temperature
profiles and the theoretical suggestions on the mean temperature profiles
consistently, we construct a model of ICM incorporating properly the
relaxation process between the ions and electrons. Fox \& Loeb (1997)
was the first to investigate the two-temperature nature of ICM. They
construct the electron and ion temperature profiles by combining the
self-similar solution by Bertschinger (1985) with the analytical
evolutionary model of adiabatic two-temperature plasma which is
originally found by Shafranov (1957). Chi\'eze, Alimi \& Teyssier (1998) 
carried out the 3-D hydrodynamical simulations of two-temperature ICM.
However, CGs only in the Einstein de Sitter universe were considered in
the previous work concerning the two-temperature model of ICM. 
Since in a higher density universe CG form at more recent epoch
(Richstone, Loeb, \& Turner 1992), the previous work is restricted to
the case in which a temperature difference is expected to be the largest
among reasonable cosmological models. Therefore, it is necessary to
study a two-temperature model of ICM in other cosmological models in
order to confirm whether the temperature difference is really
significant or not. This problem is also related to the dependence of
temperature profiles on the cosmological parameters, which is discussed 
in some numerical simulations (Evrard et al. 1996; Eke et al. 1997). 
Furthermore, quantification of resultant temperature profiles 
by using polytropic indices is not fully discussed in the previous work, 
which is very important to compare the models with X-ray observations 
properly. 
Ettori \& Fabian (1998) studied two-temperature ICM using simple analytic
models and applied their results to the CG A2163. However they neglected
the dynamical properties of ICM and assumed that all the ICM in CG is heated
at the same time. This assumption is problematic to quantify the temperature 
profiles because the timescale of shock propagation in CG is comparable to
the dynamical timescale and not much shorter than the age of the universe.

For these purpose, we apply the method of Fox \& Loeb (1997) to the
results of numerical simulations instead of the self-similar 
solution. Thus we can investigate two-temperature ICM quantitatively 
in cosmological models other than the Einstein de Sitter model. Studying
the dependence of electron temperature profiles on cosmological models
is another important purpose.

The rest of this paper is organized as follows. In \S \ref{oe}
we estimate the relevant timescales and spatial scales of ICM. In \S
\ref{eattp} we briefly review the adiabatic model of two-temperature 
plasma proposed by Fox \& Loeb (1997). In \S \ref{method} we described 
the method to calculate the ion and electron temperature profiles of ICM 
by combining a one-temperature ICM model with the adiabatic model by
Fox \& Loeb (1997). In \S \ref{simulation} we describe the adopted 
numerical methods and initial conditions for simulations of CG. 
In \S \ref{results} we present the results. In \S \ref{sd}
we summarize the results and discuss their implications.

\section{ORDER ESTIMATION}\label{oe}
We consider fully ionized plasma which consists of electrons
and ions. Only coulomb coupling is considered as the relaxation process.
Then two-body relaxation timescale of a x-particle, whose density is 
$n_{\rm x}$ and temperature is $T_{\rm x}$, is (Spitzer 1962)
\begin{eqnarray}
  t_{\rm xx} = \frac{m_{\rm x}^{1/2}(3 k T_{\rm x})^{3/2}}
                    {5.71 \pi n_{\rm x} e^4 Z_{\rm x}^4 \ln \Lambda},
  \label{eq:txx}
\end{eqnarray}
where, $m_{\rm x}$ is the particle mass, $Z_{\rm x}$ is the particle
charge number, $e$ is the electron charge, $k$ is the Boltzmann constant,
and $\ln \Lambda$ is the Coulomb logarithm and approximated to be
\begin{eqnarray}
  \ln \Lambda \simeq 37.8 + 
  \ln \biggr( \frac{T_{\rm x}}{10^8 {\rm K}} \biggl) - 
  \frac{1}{2} \ln \biggr( \frac{n_{\rm x}}{10^{-3} {\rm cm}^{-3}} \biggl)
  \label{eq:cllog}
\end{eqnarray}
for $T_{\rm x} > 4 \times 10^5$ K.
Therefore, $t_{\rm ii}$ is larger than $t_{\rm ee}$ by a factor of the
order of $(m_{\rm i}/m_{\rm e})^{1/2}$.

On the other hand, the equilibrium timescale between ions and electrons
is,
\begin{eqnarray}
  t_{\rm ei} = \frac{3 m_{\rm e} m_{\rm i}}
               {8 (2 \pi)^{1/2} n_{\rm i} Z_{\rm i}^2 e^4 \ln \Lambda}
  \biggr( \frac{kT_{\rm e}}{m_{\rm e}} + 
  \frac{kT_{\rm i}}{m_{\rm i}} \biggl)^{3/2}.
  \label{eq:tei}
\end{eqnarray}  
Therefore, $t_{\rm ei}$ is grater than $t_{\rm ii}$ by a factor of
$(m_{\rm i}/m_{\rm e})^{1/2}$. In addition, this timescale can be
comparable to or longer than the Hubble time in the outer region of CGs,
\begin{eqnarray}
  t_{\rm ei} = 2.0 \times 10^8 {\rm yr} 
           \biggr( \frac{\ln \Lambda}{40} \biggl)^{-1}
	   \biggr( \frac{n_{\rm i}}{10^{-3} {\rm cm}^{-3}} \biggl)^{-1}
           \biggr( \frac{T_{\rm e}}{10^8 {\rm K}} \biggl)^{3/2}.
  \label{eq:tei2}
\end{eqnarray}

Suppose that ICM is heated through the shock in accretion flow
(Cavaliere, Menci, \& Tozzi 1997). The shock primarily heats ions
because the kinetic energy of a particle is proportional to the particle
mass. In the post shock region the ions reach thermal equilibrium on a
timescale of $t_{\rm ii}$ after they are heated through the shock. Within
this time ion temperature is significantly higher than electron
one. Eventually thermal energy is transported from the ions to
the electrons through the Coulomb collisions between the ions and
electrons and $T_{\rm i}$ becomes comparable to be $T_{\rm e}$ on 
timescale of $t_{\rm ei}$. Note that $t_{\rm ei} > t_{\rm ee}, t_{\rm ii}$.

Under such circumstance, the radial length, $r_{\rm tt}$, over which the
electron temperature is significantly lower than the ion temperature can
be estimated as follows. Denote the propagation speed of the shock front
by $v_{\rm shock}$. Then, we have,
\begin{eqnarray}
  r_{\rm tt} \simeq t_{\rm ei} v_{\rm shock}.
  \label{eq:rtt1}
\end{eqnarray}	
Using the strong shock approximation and neglecting the post-shock gas
velocity compared with $v_{\rm shock}$, we find
\begin{eqnarray}
  v_{\rm shock} \simeq \frac{1}{3} v_{\rm infall},
  \label{eq:vshock}
\end{eqnarray}
where $v_{\rm infall}$ is the infalling velocity of the gas, which is
related to the post-shock gas temperature, $T$. The kinetic energy of
infalling gas is nearly equal to the thermal energy of the post shock
gas; namely
\begin{eqnarray}
  \frac{1}{2} m_{\rm p} v_{\rm infall}^2 \simeq k_{\rm B} T.
  \label{eq:virialT}
\end{eqnarray}
Therefore, using the equations (\ref{eq:tei2}), (\ref{eq:rtt1}), 
(\ref{eq:vshock}), and (\ref{eq:virialT}), we derive
\begin{eqnarray}
  r_{\rm tt} \simeq 1.1 \times 10^{-1} {\rm Mpc} 
                \biggr( \frac{\ln \Lambda}{40} \biggl)^{-1}
                \biggr( \frac{n_{\rm i}}{10^{-3} {\rm cm}^{-3}} \biggl)^{-1}
                \biggr( \frac{T_{\rm e}}{10^8 {\rm K}} \biggl)^{-2}.
  \label{eq:rtt2}
\end{eqnarray}
Importantly this is smaller by about one order of magnitude than 
the spatial scale of CGs. Note that
the higher temperature is, the wider becomes the region where the
temperature difference between ions and electrons is significant.

Another important timescale related to ICM is the radiative cooling
timescale, $t_{\rm c}$. Since thermal bremsstrahlung is the dominant cooling
process in ICM, we have
\begin{eqnarray}
  t_{\rm c} = 8.5 \times 10^{10} {\rm yr} 
       \biggr( \frac{n_{\rm e}}{10^{-3} {\rm cm}^{-3}} \biggl)^{-1} 
       \biggr( \frac{T_{\rm e}}{10^8 {\rm K}} \biggl)^{1/2}.
  \label{eq:tc}
\end{eqnarray}
From equations (\ref{eq:tei2}) and (\ref{eq:tc}) we find,
\begin{eqnarray}
  \frac{t_{\rm ei}}{t_{\rm c}} = 2.3 \times 10^{-3} 
           \biggr( \frac{\ln \Lambda}{40} \biggl)^{-1}
           \biggr( \frac{T_{\rm e}}{10^8 {\rm K}} \biggl).
  \label{eq:teitc}
\end{eqnarray}
Hence $t_{\rm c}$ is always longer than $t_{\rm ei}$ in the typical ICM,
and thus we can safely neglect cooling effects as far as we are concerned 
with the overall cluster structure.

On the other hand, the age of CG, $t_{\rm age}$, is of the order of 
$10^9$ or $10^{10}$ yr. Therefore, as long as heating from galaxies 
can be neglected, we can divide intra-cluster space into three regions 
according to the magnitudes of these three timescales, $t_{\rm ei}$,
$t_{\rm c}$, and $t_{\rm age}$ as follows:

\begin{enumerate}
  \item The central higher density region where 
        $t_{\rm ei} < t_{\rm c} < t_{\rm age}$. Radiative cooling is
	important but ICM can be regarded as one-temperature fluid. This
	situation corresponds to the so-cold 'cooling flow'.
  \item The middle region where $t_{\rm ei} < t_{\rm age} < t_{\rm c}$. 
        We can regard ICM as adiabatic one-temperature fluid.
  \item The outer lower density region where 
	$t_{\rm age}<t_{\rm ei}<t_{\rm c}$. Radiative cooling can be 
        negligible and the electron temperature is considerably lower
	than the ion temperature.
\end{enumerate}

\section{EVOLUTION OF ADIABATIC TWO-TEMPERATURE PLASMA}\label{eattp}

The formulation here is based on Fox \& Loeb(1997). We neglect thermal 
conduction, which is the case if tangled magnetic fields suppress conduction.
The Laglangean time evolution of the electron temperature, $T_{\rm e}$, 
and the mean temperature, 
$\bar{T}=(n_{\rm e} T_{\rm e} + n_{\rm i} T_{\rm i})/(n_{\rm e} + n_{\rm i})$,
in the adiabatic fluid element is, 
\begin{eqnarray}
  \frac{d T_{\rm e}}{dt} &=& \frac{T_{\rm i}-T_{\rm e}}{t_{\rm ei}}
  + (\gamma-1) \frac{T_{\rm e}}{n}\frac{dn}{dt}, \\
  \label{eq:dtedt}
  \frac{d \bar{T}}{dt} &=& (\gamma-1) \frac{\bar{T}}{n} \frac{dn}{dt},
  \label{eq:dtbardt}
\end{eqnarray}
where $T_{\rm i}$ is the ion temperature, $n$ is the gas density, and 
$\gamma = 5/3$, is the ratio of specific heat. 
Introducing the temperatures normalized by $\bar{T}$, 
$\tilde{T_{\rm e}} \equiv (T_{\rm e}/\bar{T})$ and $\tilde{T_{\rm i}}
\equiv (T_{\rm i}/\bar{T})$, we find,
\begin{eqnarray}
  \frac{d \tilde{T_{\rm e}}}{dt} = 
  \frac{ \tilde{T_{\rm i}} - \tilde{T_{\rm e}}}{t_{\rm ei}}.
  \label{eq:tetil1}
\end{eqnarray}
Note that $t_{\rm ei}$ is proportional to 
$\tilde{T_{\rm e}}^{3/2} \ln \Lambda$,
since in the post-shock region the gas behaves adiabatically 
($\bar{T} \propto n^{2/3}$). Thus 
$t_{\rm 2s} \equiv t_{\rm ei} (t) \tilde{T_{\rm e}} (t) ^{-3/2}$ is
constant in time, if we neglect the small change due to the Coulomb 
logarithm. Now equation (\ref{eq:tetil1}) becomes
\begin{eqnarray}
  \frac{d \tilde{T_{\rm e}}}{dt} = \frac{1}{t_{\rm 2s}} 
  \biggl( \frac{ n_{\rm i} + n_{\rm e} }{n_{\rm i}} \biggr)
  (1-\tilde{T_{\rm e}}) \tilde{T_{\rm e}}^{-3/2}.
  \label{eq:tetil2}
\end{eqnarray}
Since ICM is almost perfectly ionized, the ratio of 
$(n_{\rm i}+n_{\rm e})/n_{\rm i}$ can be regarded as constant.
Thus equation (\ref{eq:tetil2}) can be integrated analytically. If
we assume the pre-shock $\tilde{T_{\rm e}}$ equal to zero, the solution
for the fluid element which has passed the shock front at $t=t_{\rm s}$ is,
\begin{eqnarray}
  t-t_{\rm s} = t_{\rm 2s} \biggl( \frac{n_{\rm i}}{n_{\rm i}+n_{\rm e}} 
                \biggr)
            \biggl[
            \ln \biggr( 
            \frac{1+\sqrt{\tilde{T_{\rm e}}}}{1-\sqrt{\tilde{T_{\rm e}}}}
            \biggl)
            -2 \sqrt{\tilde{T_{\rm e}}}
            \biggr( 1 + \frac{\tilde{T_{\rm e}}}{3} \biggl)
            \biggr].
   \label{eq:ttetil}
\end{eqnarray}
Thus we can obtain the Laglangean time evolution of $\tilde{T}_{\rm e}$ 
of the fluid element after the passage through the shock by solving 
equation (\ref{eq:ttetil}).

\section{NUMERICAL METHOD}\label{method}
Using equation (\ref{eq:ttetil}) we can construct the temperature profiles 
at $t=t_0$, $T_{\rm e}(r,t_0)$ and $T_{\rm i}(r,t_0)$, as follows. First, 
we give the velocity field of the gas, $v(r,t)$, the shock radius, 
$r_{\rm shock}(t)$, and the equilibrium timescale of the fluid element 
at the passage through the shock front, $t_{\rm 2s}[r_{\rm shock}(t)]$,
using some analytical models or numerical simulations presented later.
Then we can calculate the Laglangean path of the fluid element which
passed the shock surface at $t=t_{\rm s}$, $R(t;t_{\rm s})$, by solving
the ordinary differential equation,
\begin{eqnarray}
  \frac{dR}{dt} &=& v( R(t;t_{\rm s}),t ),
  \label{eq:drdt}
\end{eqnarray}
with the appropriate initial condition,
\begin{eqnarray}
  R(t_{\rm s};t_{\rm s}) &=& r_{\rm shock}(t_{\rm s}).
  \label{eq:rt0}
\end{eqnarray}
Integrating differential equation (\ref{eq:drdt}) for various 
$t_{\rm s}$, we obtain $r_0=R(t_0;t_{\rm s})$, the radius which the fluid
element that passed the shock at $t=t_{\rm s}$ resides at $t=t_0$, as a
function of $t_{\rm s}$. Thus we can regard $t_0-t_{\rm s}$ as a
function of the radius,
\begin{eqnarray}
  t_0-t_{\rm s} = f(r;t_0).
  \label{eq:t0tsr}
\end{eqnarray}
Using the equations (\ref{eq:ttetil}) and (\ref{eq:t0tsr}), and 
the model of $t_{\rm 2s}[r_{\rm shock}(t_{\rm s})]$, we can solve
$\tilde{T_{\rm e}}$ at $t=t_0$ as a function of $r$. Finally, we obtain
$T_{\rm e}(r)$ and $T_{\rm i}(r)$ using $\tilde{T}_{\rm e}(r)$ and the
model of $\bar{T}(r)$.

\section{THE SIMULATIONS}\label{simulation}

\subsection{Numerical Method}
To give $v(r,t)$, $r_{\rm shock}(t)$, etc, we have performed
numerical simulations of a spherically symmetric CG. For dark matter (DM),
we use the shell model (H\'{e}non 1964).
We set the number of shells, $N$, equal to 5000.
As for gas on the other hand, we use 1-dimensional,
spherically symmetric, total variation diminishing (TVD) code with minmod
limiter (Hirsch 1990). Note that TVD code is one of the most powerful 
tools to treat shocks. One mesh spacing corresponds to 
$\Delta r = 0.005/(1+z)$ Mpc. We assume that the gas is ideal, with 
$\gamma = 5/3$. As to the boundary conditions, the inner edge is 
assumed to be a perfectly reflecting point. The outer edge is assumed 
to be a perfectly transmitting surface.
The basic equations and the numerical method used here are fully described 
in \S 2 of Takizawa \& Mineshige (1998).

\subsection{Models and Initial Conditions}
In this paper, all of the calculations are carried out from $z_{\rm ini}=10$
to the present time ($z_0=0$). 
The cosmological models which we adopt are
$(\Omega_0, \Lambda_0)=(1.0, 0.0)$ (ED), 
$(\Omega_0, \Lambda_0)=(0.2, 0.0)$ (OP), and
$(\Omega_0, \Lambda_0)=(0.2, 0.8)$ (FL). 
The Hubble constant, $H_0$ is set to be 
$H_0=100 {\rm km}{\rm s}^{-1}{\rm Mpc}^{-1}$ in the transformations of 
length and time coordinates. 
We set $\Omega_{\rm b}=0.012 h^{-2}$ in all models taken from the
nucleosynthesis determination (Walker et al. 1991).
Note that $\Omega_{\rm DM} = \Omega_0 - \Omega_{\rm b}$.

We make initial density profiles in the same manner as Peebles (1982).
At first we prepare $N$ concentric shells with a constant density equal to
$\Omega_{\rm DM}$ at $z_{\rm ini}=10$. Then,
a density fluctuation is introduced by perturbing the radius and velocity
of each shell following
\begin{eqnarray}
  r_i &=& r_i^{(0)} \biggr\{ 1 - 
         \frac{1}{3} \bar{\delta}[ r_i^{(0)} ] \biggl\},  \\
  \label{eq:ri}
  u_i &=& H(z_{\rm ini}) r_i^{(0)} \biggr\{ 1 - 
         \frac{1}{3}\bigr(1+\Omega^{0.6}\bigl)
                                       \bar{\delta}[ r_i^{(0)} ] \biggl\},
  \label{eq:ui}
\end{eqnarray}
where $r_i^{(0)}$ is the unperturbed coordinate, $\bar{\delta}(r)$ 
represents the mean density fluctuation inside $r$, which is derived 
from the density fluctuation field (specified below) within $r$, and
$H(z_{\rm ini})$ is the Hubble constant at $z=z_{\rm ini}$. Here,
we used the Zel'dovich approximation (Zel'dovich 1970) and 
the approximation of $d \log D_1 / d \log a \simeq \Omega^{0.6}$,
where $D_1$ is the linear perturbation growth rate for the growing mode
and $a$ is the scale factor (see Suto 1993).

The initial conditions of the density fluctuation field, $\delta(r)$,
are generated by applying the Hoffman-Ribak method (Hoffman \& Ribak
1991; van de Weygaert \& Bertschinger 1996) to spherical systems (see
Appendix). We constrained in such a way that there exist initially
density enhancements on 1 Mpc scale whose amplitudes correspond to 
$3 \sigma$ level in the CDM power spectrum. The normalization is $\sigma_8=1$ 
in each cosmological model, which was obtained from the observation 
of the nearby galaxy distribution (see Suto 1993).

The initial conditions of gas are set as follows. At first, gas density
was everywhere taken to be the mean baryon density of the universe at
$z=10$, and the temperature of the gas ($T_{\rm gas, i}$) was constant
everywhere; $T_{\rm gas, i} = 10^7 {\rm K}$ in all models. We then add
the adiabatic fluctuation in such a way that the ratio of the DM density
and the gas density remains the same. Note that after the perturbation
is added, the gas temperature distribution becomes nonuniform,
accordingly. Moreover, note that temperature of the infalling gas at 
$z\sim 1$ is sufficiently lower than virial temperature, since the gas
expands adiabatically following the cosmological expansion.

\section{RESULTS}\label{results}

The overall evolution of the simulated CG is essentially the same as
that presented by Takizawa \& Mineshige (1998). Thus we concentrate
on showing the results of the temperature profiles, mass estimation,
and so on.

\subsection{Temperature Profiles}
Figure \ref{fig:temed} shows radial distribution the $T_{\rm e}(r)$ 
(solid line), $T_{\rm i}(r)$ (dotted line), and $\bar{T}(r)$ 
(short dashed line) of Model ED at $z=0$. At $r<0.6$Mpc $T_{\rm e}$
is very close to $T_{\rm i}$, while, at
$r>0.6$Mpc the electron temperature is considerably lower than the ion
temperature and the discrepancy gets increased outward. At $r \simeq
1$Mpc, the electron temperature is only a half of the mean temperature.
Small scale fluctuations in the temperature profiles are due to the sound
wave propagation in ICM (Takizawa \& Mineshige 1988).

In real X-ray observations what is actually obtained is an 
emissivity-weighted, line-of-sight projected electron temperature map,
which is displayed in Figure (\ref{fig:temobs}) for model ED. 
In this figure, we assumed the spatial resolution to be 0.25 Mpc, 
which corresponds to about 6' for an object located at $z=0.05$,
and the error of the temperature measurement to be 20 \%.
Also in this map we can clearly see that the electron temperature 
decreases outward.

To estimate the temperature gradients both for $T_{\rm e}$ and
$\bar{T}$, we measure the polytropic indexes, $\gamma_{\rm p}$,
in the usual way. We fit the density profile by the $\beta$-model 
and the temperature profile by the polytropic model as follows,
\begin{eqnarray}
  n(r) &=& n_0 \biggr[ 1 + \biggr( \frac{r}{r_{\rm c}} \biggl)^2 \biggl]
        ^{-3 \beta/2},
  \label{eq:betamodel} \\
  T(r) &\propto& n(r)^{\gamma_{\rm p}-1}
  \label{eq:tprof}
\end{eqnarray}
where $n_0$, $r_{\rm c}$, and $\beta$ is the fitting parameters of the 
$\beta$-model. We fit the resultant density profile only inside the 
shock front. The data are fitted by chi-square fitting. We assume 
that the variance in any quantity is proportional to its square 
($\delta f/f = {\rm const.}$) because our main purpose is not
to simulate observations with a specific instrument but to obtain
the intrinsic profiles of our calculated results. 
The results of the fitting are summarized in 
Table \ref{tab:fpden} and the corresponding polytropic indices are listed
in Table \ref{tab:pi}. From Table \ref{tab:fpden} we find that the
$\beta$ values are somewhat bigger than those typically observed. On the
other hand, our results are well coincident with the self-similar
solution by Bertschinger (1985). Thus the bigger $\beta$ values are
probably due to the assumption of spherical symmetry and neglect of
angular momentum of gas. If the initial gas temperature is higher,
which corresponds to the case that reheating of the ICM, e.g., by 
protogalaxies, is substantial, $\beta$ values can be smaller
(Metzler \& Evrard 1994; Takizawa \& Mineshige 1988).
From Table \ref{tab:pi} we find that the
polytropic indices of the electron temperature profiles are 
$\gamma_{\rm p} \sim 1.5$, which are systematically larger than those of
the mean temperature profiles, $\gamma_{\rm p} \sim 1.3$.

The specific entropy profiles derived from the electron temperature
$[S_{\rm e} \propto \ln(T_{\rm e}/\rho^{\gamma-1})]$, and from the
mean temperature $[\bar{S} \propto \ln(\bar{T}/\rho^{\gamma-1})]$, 
are shown in Figure \ref{fig:sesbar} by the solid and dotted lines,
respectively. Entropy is normalized to be zero at the inner boundary. 
The latter $\bar{S}$ rises outward, which is characteristic of the ICM 
heated through the shock (Evrard 1990; Takizawa \& Mineshige 1998). 
The former $S_{\rm e}$, in contrast, rises outward as the case only
in the inner region ($r<0.6$Mpc), stays nearly constant in the middle 
region, and falls outward in the outer region.

\subsection{Dependence of Fitting Parameters on Size of Fitting}

We fit the calculated density profile inside the shock front by equation 
(\ref{eq:betamodel}) and listed the fitting results in Tables 
\ref{tab:fpden} and \ref{tab:pi}. However, it is possible that the fitting 
results can be influenced by the position of the outer edge of the region
used for the fitting. In real X-ray observations the outer edge of 
the X-ray emitting region is perhaps inside the shock front. 
To assess this effect we fit the density and temperature profiles 
inside various radii for Model ED and list the dependence of the fitting 
results related to the density profile ($n_0$, $r_{\rm c}$, and $\beta$)
on the outer-edge radius ($r_{\rm out}$) in Figure \ref{fig:fparaden}.
It is found that these parameters are insensitive to the outer
radius as long as $r_{\rm out}>0.8$ Mpc. In density profiles, therefore,
we can safely neglect the influence of the outer edge. This fact is 
actually expected because of the self-similar nature of the gas density 
profile (Bertschinger 1985; Takizawa \& Mineshige 1998).

On the other hand, the dependence of $\gamma_{\rm p}$ on $r_{\rm out}$ 
is rather different. Figure \ref{fig:fparatem} shows that $\gamma_{\rm p}$
derived from $T_{\rm e}$ (by closed squares) monotonically increase outward
whereas $\gamma_{\rm p}$ derived from $\bar{T}$ (by open square) does not
exhibit systematic changes. When we calculate a polytropic index from 
X-ray observational data, thus, the resultant value could be subjected
to large errors arising from the finite detection limit, background noise, 
and so on.

The results of model OP and FL are essentially the same as those of  model ED.
Although there is a simple self-similar solution only for model ED,
self-similar nature is also expected both in OP and FL
(Takizawa \& Mineshige 1988).

\subsection{Mass Estimation}
Since the observed electron temperature significantly deviates from the
mean temperature that determines the dynamics of the system,
the total mass of CG is probably underestimated if hydrostatic
equilibrium is calculated based on the electron temperature map.
When ICM is assumed to be isothermal we usually use the emissivity-weighted 
mean temperature, which is more like the temperature in the central
high-density region. Thus, the underestimation of the mass is practically
negligible. When the electron temperature profile is used, conversely,
the mass can be seriously underestimated. Hence the mass derived from 
the assumption of hydrostatic equilibrium, $M_{\rm hydro}(r)$, is,
\begin{eqnarray}
  M_{\rm hydro} (r) = \frac{3 k T_0 \beta \gamma_{\rm p}}{\mu m_{\rm p} G} r
  \frac{ (r/r_{\rm c})^2 }{ [ 1 + (r/r_{\rm c})^2]^{1+3 \beta (\gamma_{\rm p}-1)/2}  }.
  \label{eq:mhse}
\end{eqnarray}
Figure \ref{fig:mass} depicts the ratio of $M_{\rm hydro}$ to the actual mass
as a function of radius. The solid line represents the ratio calculated 
based on $T_{\rm e}(r)$ and the dotted line represents that based on 
$\bar{T}(r)$. The mass derived from $T_{\rm e}$ is underestimated 
by almost 50 \% because of the lower electron temperature.
Note that the mass derived from $\bar{T}$ is also underestimated by
about 10 \% due to the bulk motion of the gas.

In the central region $M_{\rm hydro}$ based on $T_e$ and $\bar{T}$ are 
both larger than the actual mass. This is due to the fitting
errors of temperature profiles. Since the core radii are different between 
the density and temperature profiles, polytropic model cannot describe
the our results well in this region.

\section{SUMMARY AND DISCUSSION}\label{sd}
We constructed the models of ICM, incorporating the relaxation process 
between the ions and electrons. From the simple order estimation, we find
that the electron temperature is well below the ion temperature 
in the outer region of CG and that such a lower $T_{\rm e}$ region spreads
over a Mpc scale in typical CG. In addition, the hotter CG is, the wider
becomes the two-temperature region. Comparing three relevant timescales 
in ICM (the age of CG, radiative cooling, and equilibrium timescales
 between ions and electrons), we can divide ICM into three regions; 
from the center outward, the cooling dominant, one-temperature region, 
adiabatic, one-temperature region, and adiabatic, two-temperature region.

We calculate the temperature profiles of two-temperature ICM combining 
the spherically symmetric, N-body and hydrodynamic simulations for three 
different cosmological models with the adiabatic two-temperature 
plasma model by Fox \& Loeb (1997). While the polytropic indices of the 
mean temperature profiles are $\simeq 1.3$, those of the electron 
temperature profiles are $ \simeq 1.5$. As a consequence, the specific 
entropy profiles derived from the electron temperature are rather flat.

We examine the dependence of the fitting parameters on the radius of the
outer edge of the region used for the fitting when the density profile is 
fitted by the $\beta$-model and the temperature profile by the polytropic 
model. The fitting results of the density profile and mean temperature 
profile is insensitive to the outer edge. On the other hand, the polytropic 
index derived from the electron temperature rises as the outer-edge radius 
increases.

The total mass of CG is underestimated about 50 \% when 
we use the electron temperature profile though the underestimation is
negligible when we assume that ICM is isothermal and adopt 
emission-weighted mean temperature as the temperature of ICM. 

We confirm that temperature difference between ions and electrons in ICM is
substantial in cosmological models other than the Einstein de Sitter model.
The polytropic indices of electron temperature profiles are insensitive
to the cosmological models in the range of our calculations.  
Note that the baryon density of these models is set to be the same. 
If the baryon fraction is set to be constant 
in each model, the result will probably changes because the equilibrium 
timescale is sensitive to the baryon density.

In general our resultant electron-temperature profiles tend to have steeper
gradients than those of Ettori \& Fabian (1998). They assumed that all the 
ICM in CG is heated at the same time and that ICM is in hydrostatic 
equilibrium. On the other hand, in our calculations the ICM in the outer 
region is heated more recently than that in the inner region because 
the shocks propagate at finite speed. Furthermore, gradual radial infall 
persists inside the shock fronts and
makes the two-temperature region slightly compress inward.
(Takizawa \& Mineshige 1998).   
Thus the two-temperature region of our results becomes
wider than those of Ettori \& Fabian (1988).

Although only spherical accretion is considered in this paper, 
there may also arise asymmetric merging of comparable clumps in reality.
It is possible that shocks occurring in merging events will also generate
the temperature difference between ions and electrons. In this case 
it is believed that a bow shock with an arc shape is formed just 
between the centers of two substructures (Schindler \& M\"{u}ller 1993;
Ishizaka \& Mineshige 1996; Roettiger et al. 1997; Ishizaka 1998). 
In the post-shock region 
the energy is transported from ions to electrons but that timescale 
changes along the shock front since the timescale is sensitive 
to the density of ICM. Therefore, if we consider the temperature
difference between ions and electrons, the location of the observed hot 
gas region is probably shifted and the shape is more deformed in comparison
with the results of the former simulations. We will investigate this issue
as a future work.

We consider only the classical coulomb coupling as the relaxation process
between ions and electrons. It is possible, however, in ICM more efficient
relaxation processes may be effective (McKee \& Cowie 1977; 
Pistinner, Levinson, \& Eichler 1996). In this case 
the equilibrium timescale can be shorter than the value given by equation
(\ref{eq:tei2}). Therefore, the temperature difference between ions and
electrons can be less and polytropic indices can be smaller than our
results. If magnetic field exits in ICM, it is possible that
electrons are also significantly heated in shocks by MHD instabilities.
Also in this case the temperature difference between ions and
electrons can be less.

We neglect the heating process from the galaxies to ICM. If
thermalized hot gas is injected to ICM from the galaxies, temperature
difference could be less. Furthermore, in this case the preheating of the 
ICM influences the density profile and mean temperature profile (Metzler \& 
Evrard 1994; Navarro et al. 1995; Takizawa \& Mineshige 1998). 
Since heavy elements like iron are detected in ICM at redshift up to 
$z \simeq 1$ (Hattori et al. 1997), this effect should be 
considered to construct a more realistic model. In such a study the
adiabatic model like Fox \& Loeb (1997) cannot be adopted. Thus
non-adiabatic models or fully two-fluid simulations like Chi\'eze et
al. (1998) are required.

\acknowledgements

The author would like to thank S. Mineshige for valuable comments. This
work is supported by a Grant-in-Aid from the Ministry of Education, 
Science, Sports and Culture of Japan (6179) and
Research Fellowships of the Japan Society for the 
Promotion of Science for Young Scientists.

\appendix
\section{THE HOFFMAN-RIBAK METHOD FOR SPHERICAL SYSTEMS}
We consider a random homogeneous and isotropic Gaussian field $f$ 
with zero mean which is defined by its power spectrum $P(k)$. 
When $f$ is subjected to linear constraints, $g = C f$, then
the constrained Gaussian field $f$ is realized as follows 
(Hoffman \& Ribak 1991),
\begin{eqnarray}
  f = \tilde{f} + M C^{\dag}Q^{-1}(g - C \tilde{f}),
\end{eqnarray}
where $\tilde{f}$ is an unconstrained random Gaussian field whose power
spectrum is $P(k)$, $M$ is the two-point correlation matrix obtained from
$P(k)$, and $Q=CMC^{\dag}$.

In general three dimensional case, $\tilde{f}$ is described by its Fourier
components $\tilde{f}_{\bf k}$,
\begin{eqnarray}
  \tilde{f}({\bf r}) = \frac{1}{(2 \pi)^3} \int \tilde{f}_{\bf k} 
                       \exp(i {\bf k}{\bf r}) d{\bf k}.
\end{eqnarray}

However, to generating the initial conditions which can be used 
for our spherical symmetric simulations, we should construct 
$\tilde{f}$ as a function of only $r$ instead of ${\bf r}$. 
Since $\tilde{f}$ is isotropic $\tilde{f}(r)$ is obtained as follows.
In spherical systems it is convenient to expand 
$\tilde{f}({\bf r})$ in spherical harmonics times spherical Bessel 
functions $j_l$ (Binney \& Quinn 1991). A plain wave is expanded 
in these functions as follows,
\begin{eqnarray}
  \exp(i {\bf k}{\bf r}) = 4 \pi \sum_{l=0}^{+\infty} \sum_{m=-l}^{+l}
  i^l j_l(kr) Y_l^{m*} (\theta_k, \phi_k) Y_l^m (\theta_r, \phi_r).
\end{eqnarray}
Integrating this for the tangential components we find,
\begin{eqnarray}
  \int \exp( i {\bf k}{\bf r} ) d {\bf k} = 
  4 \pi \int j_0(kr) k^2 dk.
\end{eqnarray}
We can assume that $\tilde{f}_{\bf k} = \tilde{f}_k$ because of isotropy of
$\tilde{f}$. Thus we obtain,
\begin{eqnarray}
  \tilde{f}(r) = \frac{1}{2 \pi^2} \int \tilde{f}_k \frac{\sin kr}{kr} k^2 dk.
\end{eqnarray}

\clearpage
 \begin{table}
  \begin{center}
   \begin{tabular}{cccc} 
    \hline \hline
    Model  & $n_0$ ($10^{-3}$cm$^{-3}$) & $r_{\rm c}$ ({\rm Mpc}) &  $\beta$ \\ 
    \hline
    ED      & $1.16$      & $0.067$         &  $0.793$        \\
    OP      & $3.64$      & $0.104$         &  $0.815$        \\
    FL      & $1.45$      & $0.155$         &  $0.903$        \\
    \hline
   \end{tabular}
  \end{center}
  \caption[Density profiles of gas at $z=0$.]{Density profiles of gas at $z=0$.}
  \label{tab:fpden}
 \end{table}%

\clearpage
 \begin{table}
  \begin{center}
   \begin{tabular}{ccc} 
    \hline \hline
    Model  & $\gamma_{\rm p}$ of $T_{\rm e}$  & $\gamma_{\rm p}$ of $\bar{T}$   \\ 
    \hline
    ED      & $1.55$      & $1.35$   \\
    OP      & $1.44$      & $1.34$   \\
    FL      & $1.51$      & $1.37$   \\
    \hline
   \end{tabular}
  \end{center}
  \caption[Polytropic indices of $T_{\rm e}$ and $\bar{T}$ at $z=0$.]{Polytropic indices of $T_{\rm e}$ and $\bar{T}$ at $z=0$.}
  \label{tab:pi}
 \end{table}%

\newpage

\figcaption{The profiles of the $T_{\rm e}$ (solid line), $T_{\rm i}$ 
     (dotted line), and $\bar{T}$ (short dashed line) of Model ED at 
     $z=0$. At $r<0.6$ Mpc $T_{\rm e} \sim T_{\rm i}$, whereas at 
     $r>0.6$ Mpc $T_{\rm e}$ is considerably lower than $T_{\rm i}$
     and the discrepancy gets enhanced outward. At $r \simeq 1$ Mpc, 
     $T_{\rm e} \sim \bar{T}/2$. Small scale fluctuations in the temperature 
     profiles are due to the sound wave propagation in ICM.
     \label{fig:temed}}

\figcaption{Emissivity weighted, line-of-sight projected electron 
     temperature profile of Model ED at $z=0$. Here we assumed 
     the spatial resolution to be 0.25 Mpc, which corresponds to about 6' 
     for an object located at $z=0.05$, and the error of the temperature 
     measurement to be 20 \%. Clearly the electron temperature decreases 
     outward.
     \label{fig:temobs}}

\figcaption{The specific entropy profiles derived from the electron 
     temperature $[S_{\rm e} \propto \ln(T_{\rm e}/\rho^{\gamma-1})]$, 
     by the solid line and from the mean temperature 
     $[\bar{S} \propto \ln(\bar{T}/\rho^{\gamma-1})]$, by the dotted line,
     respectively. Entropy is normalized to be zero at the inner boundary.
     \label{fig:sesbar}}

\figcaption{The dependence of the parameters related to the density profile 
     ($n_0$ , $r_{\rm c}$, and $\beta$) on $r_{\rm out}$, the radius of 
     the outer edge of the region used for fitting. It is found that these 
     parameters are insensitive to the outer edge radius as long as 
     $r_{\rm out}>0.8$.
     \label{fig:fparaden}}

\figcaption{The dependence of $\gamma_{\rm p}$ on $r_{\rm out}$. We find that 
     $\gamma_{\rm p}$ obtained from $T_{\rm e}$ (closed squares) increases
     as $r_{\rm out}$ increases, whereas $\gamma_{\rm p}$ obtained from 
     $\bar{T}$ (open squares) is insensitive to $r_{\rm out}$.
     \label{fig:fparatem}}

\figcaption{The ratio of $M_{\rm hydro}$ to the actual mass. The solid 
     line is the result based on $T_{\rm e}(r)$ and the dotted line is 
     that based on $\bar{T}(r)$. The mass derived from $T_{\rm e}$ is 
     underestimated by almost 50 \% because of the lower electron 
     temperature. Note that the mass derived from $\bar{T}$ is also 
     underestimated by about 10 \% due to the bulk motion of the gas
     \label{fig:mass}}
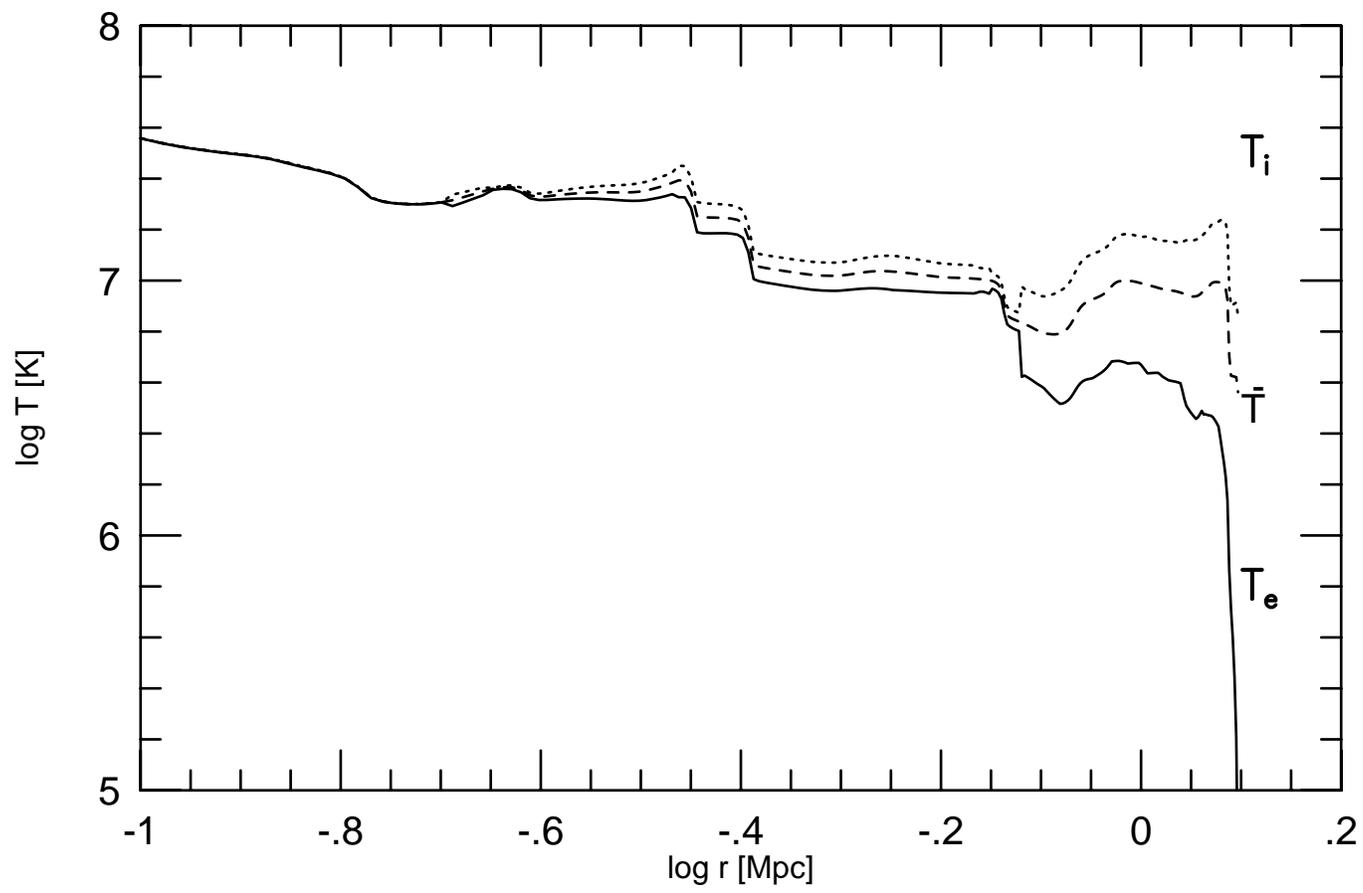
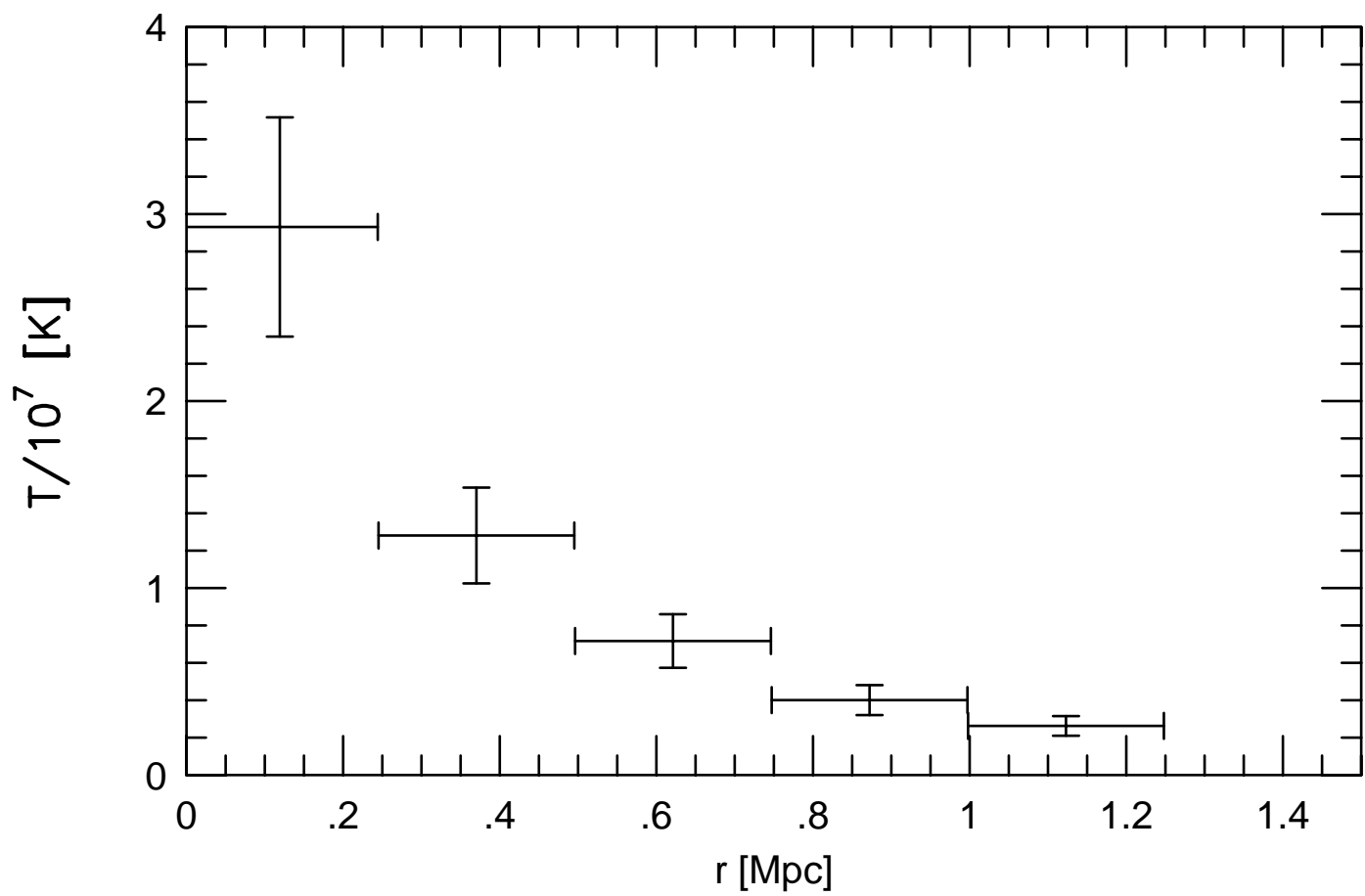
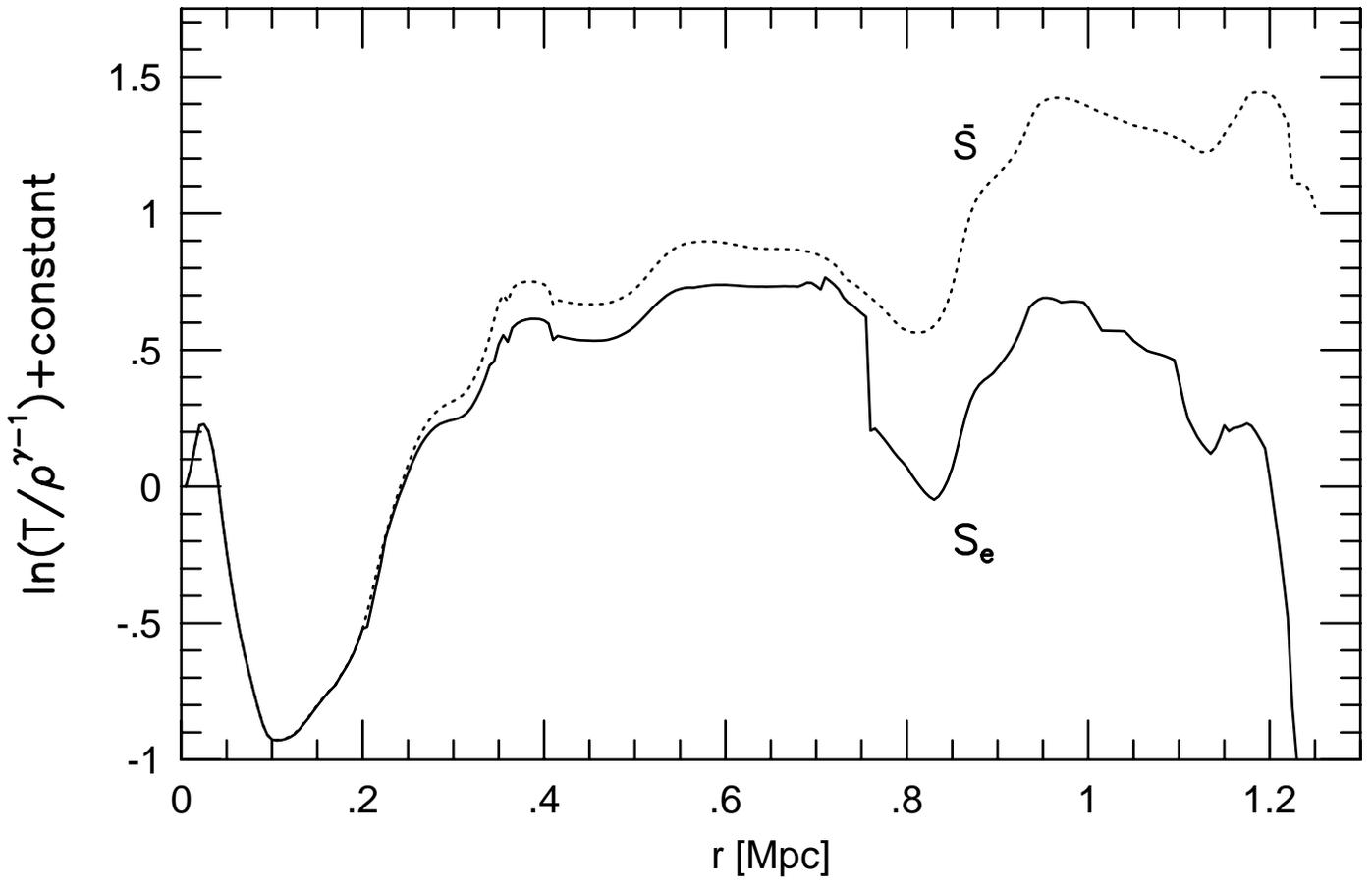
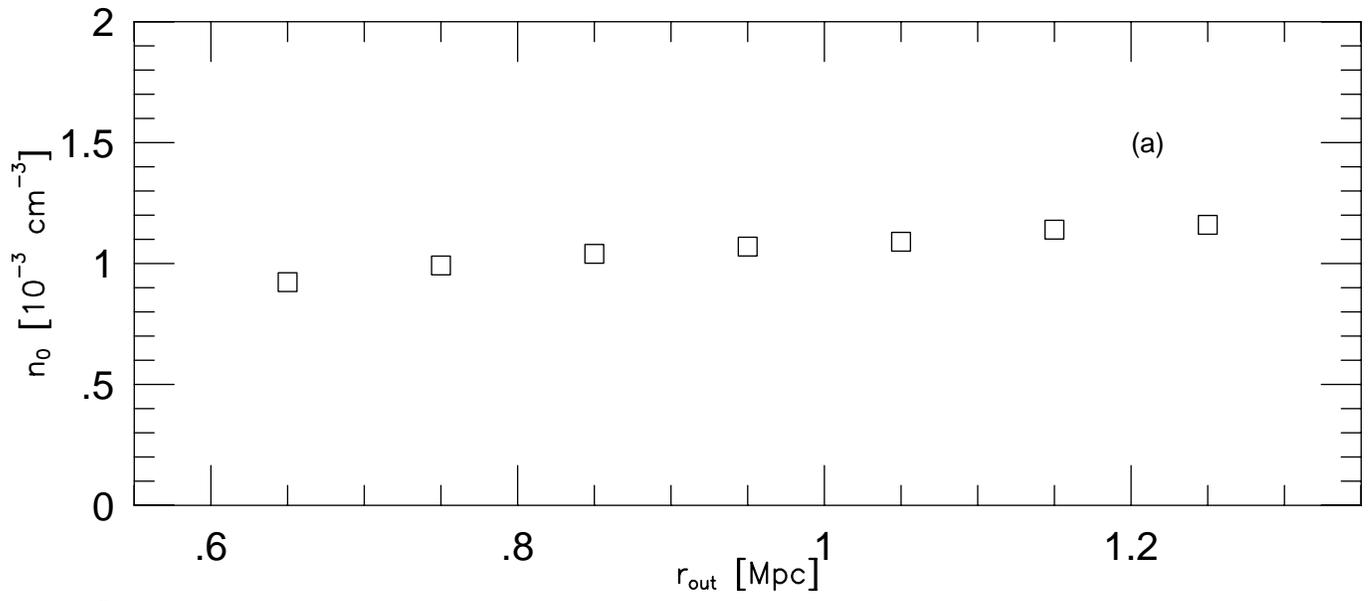
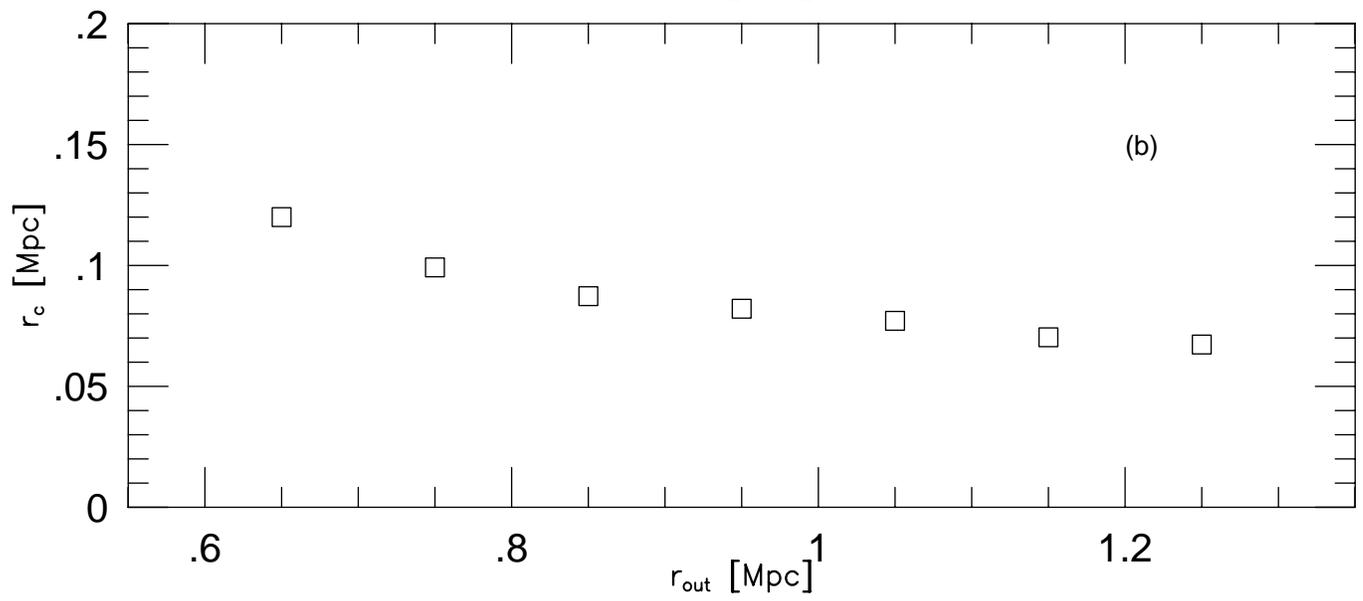
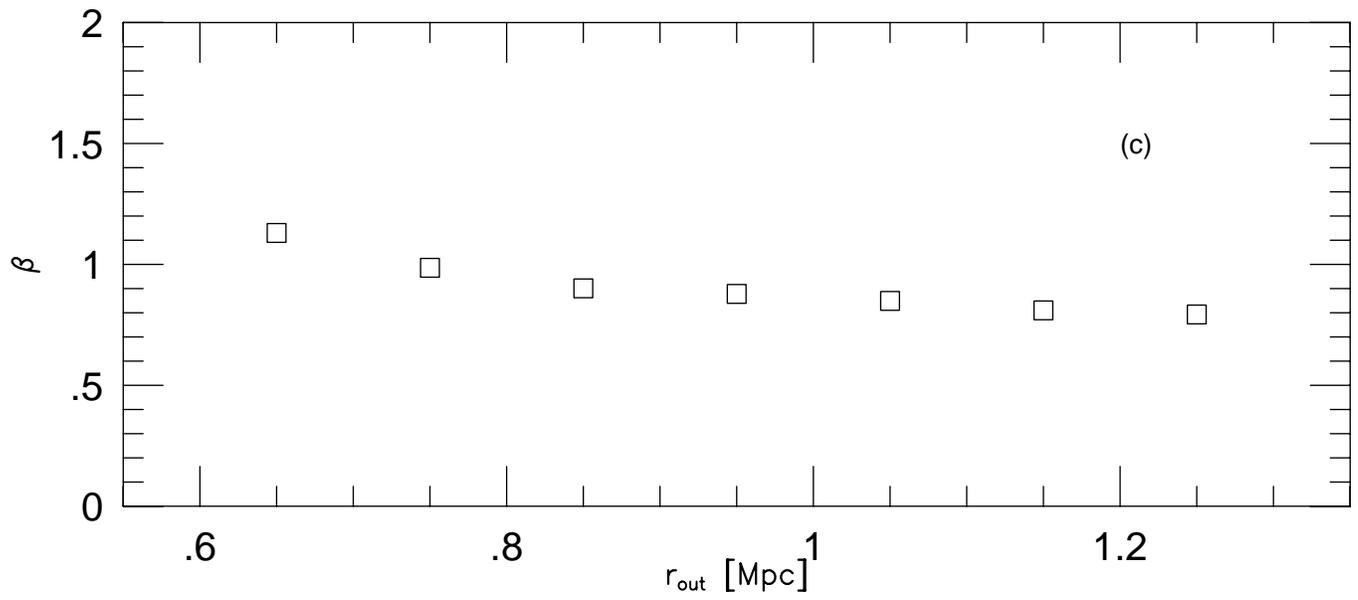
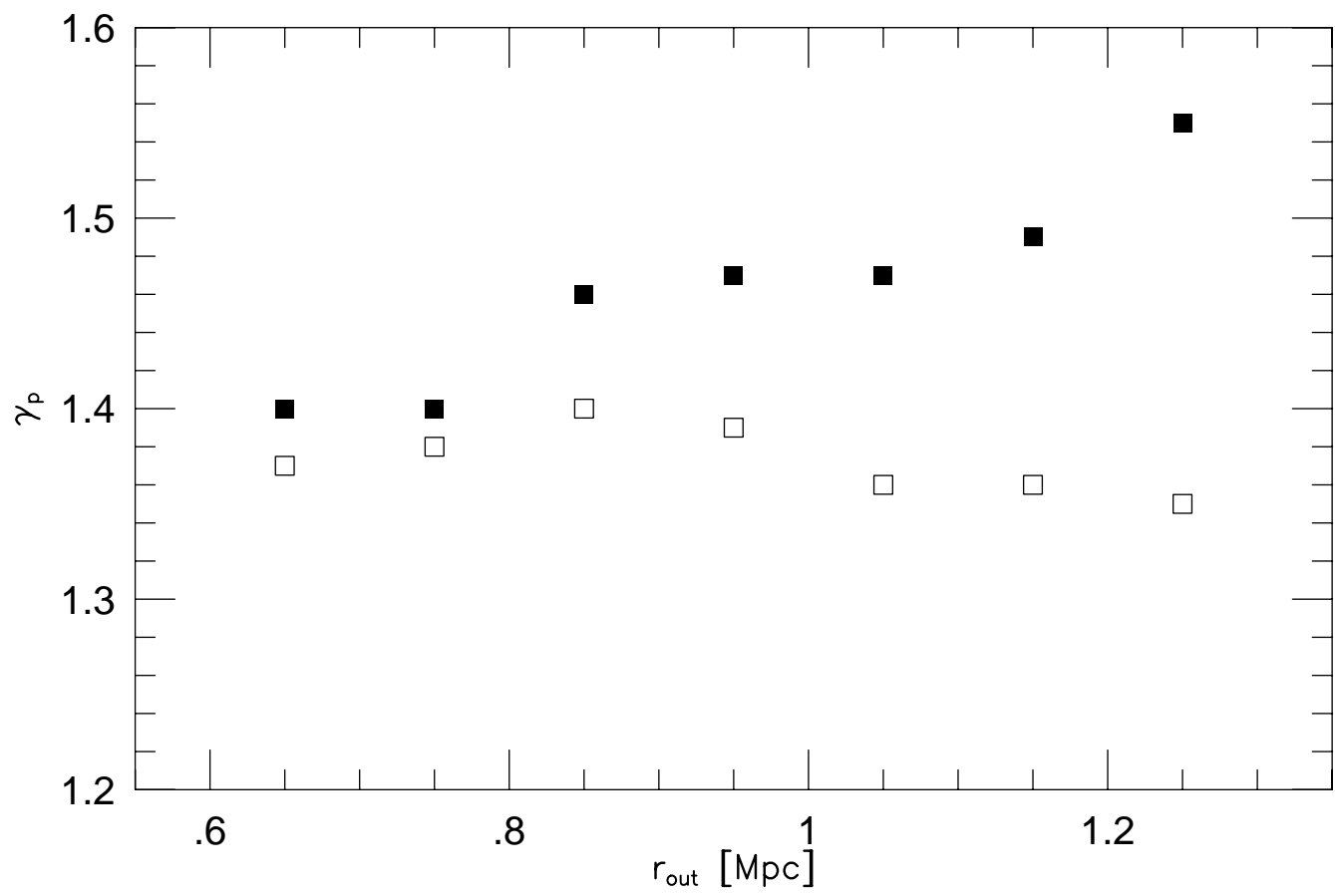
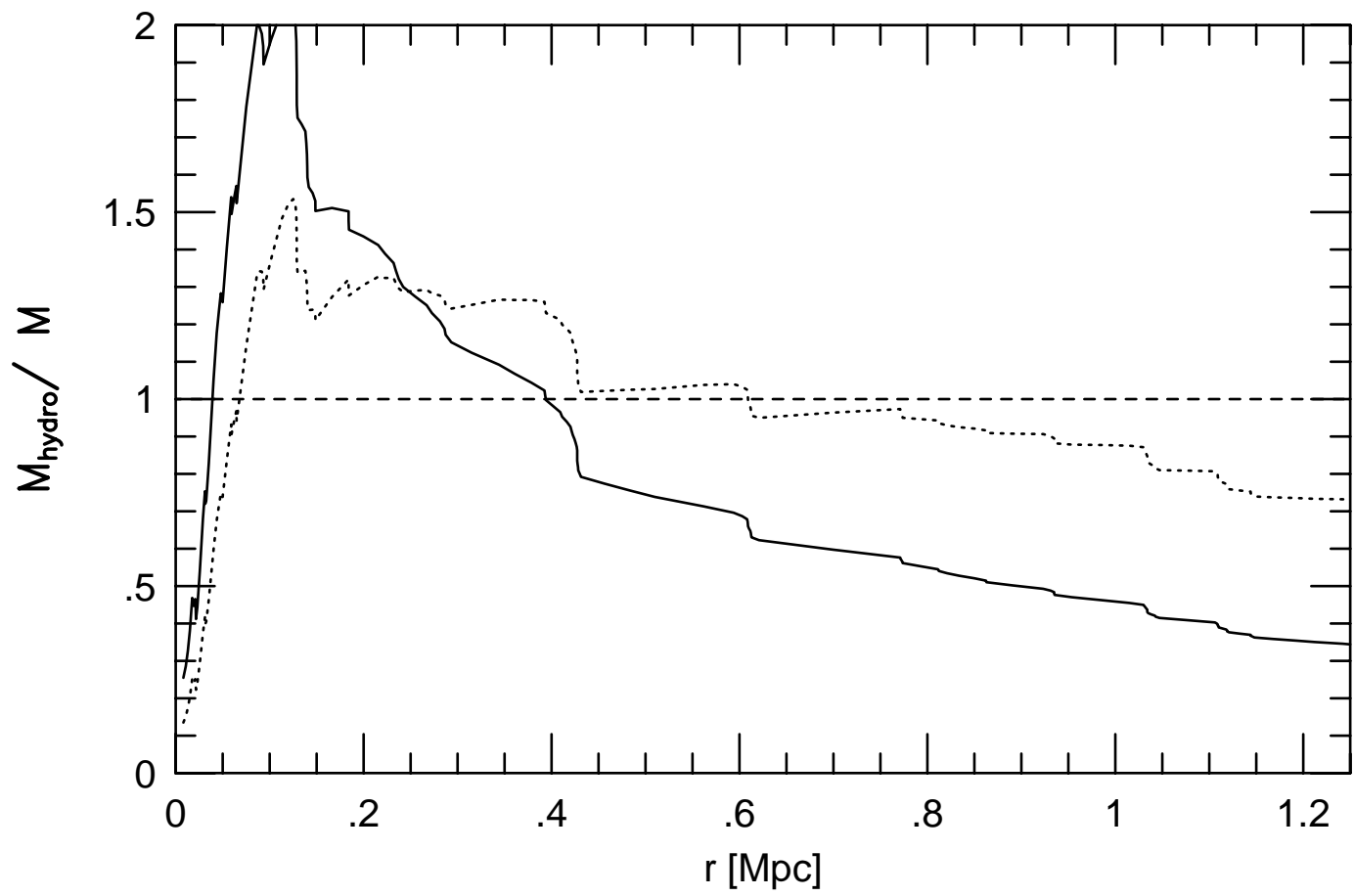
\end{document}